\documentclass[preprint,12pt]{elsarticle}

\usepackage{amsmath,amsfonts,amsthm,amssymb,paralist,subfigure,graphicx,amsbsy,float,epsfig,color}
\usepackage{cuted,mathtools,lipsum}

\usepackage[ruled,vlined,linesnumbered]{algorithm2e}

\usepackage{stmaryrd,url}

\usepackage{amsthm}

\newtheorem{defn}{Definition}
\newtheorem{rem}{Remark}

\def\mb{\mathbf}

\def\mc{\mathcal}

\def\mb{\mathbf}

\def\mc{\mathcal}

\journal{European Journal of Control}

\begin{document}

\begin{frontmatter}

\title{Distributed Observer-based Fault Detection over Intelligent Networked Multi-Vehicle Systems
}

\author[Sem]{Mohammadreza Doostmohammadian}
\affiliation[Sem]{Mechatronics Group, Faculty of Mechanical Engineering, Semnan University, Semnan, Iran, doost@semnan.ac.ir.}

\author[HR]{ Hamid R. Rabiee}
\affiliation[HR]{Computer Engineering Department, Sharif University of Technology, Tehran, Iran,
	rabiee@sharif.edu.}

\begin{abstract}
Decentralized strategies are of interest for local decision-making over multi-vehicle networks. This paper studies mixed traffic networks of human-driven and autonomous vehicles with partial sensor measurements. The idea is to enable the group of connected autonomous vehicles (CAVs) to track the state of a group of human-driven vehicles (HDVs) via distributed consensus-based observers/estimators. Particularly, we make no assumption that the group of HDVs is locally observable in the direct neighborhood of any CAV. Then, the main contribution is to design \textit{local} residual-based fault detection and isolation (FDI) at every CAV to detect possible faults/attacks in the sensor measurements. This distributed detection strategy enables every CAV to \textit{locally} find possible anomalies in its taken sensor measurement with no need for a central processing unit. Two FDI logics are proposed with and without considering the history of the residuals. These FDI techniques are based on probabilistic threshold design on the residuals (in contrast to the existing deterministic threshold FDI techniques) with no assumption that the noise is of bounded support. This is more realistic in real-world multi-vehicle transportation systems.
\end{abstract}

\begin{graphicalabstract}
	\includegraphics{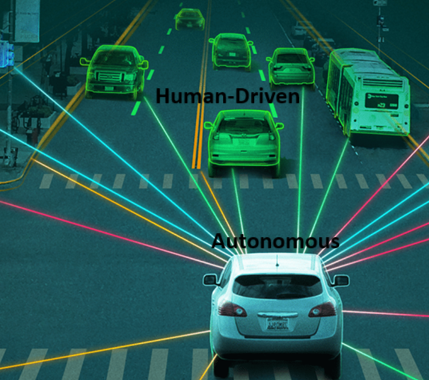}
\end{graphicalabstract}

\begin{highlights}
	\item Studying the mixed traffic network of human-driven and autonomous vehicles taking partial sensor measurements 
	\item Proposing distributed consensus-based observers to track the state of the human-driven vehicles
	\item Designing local residual-based fault detection methods to detect possible faults/attacks in the sensor measurements
	\item Proposing both stateless and stateful localized fault detection algorithms based on probabilistic threshold design
\end{highlights}


\end{frontmatter}

\section{Introduction} \label{sec_intro}
Distributed algorithms are gaining traction across various applications due to several key advantages over traditional centralized approaches, including scalability, resilience, redundant processing and collective intelligence 	\cite{2025survey}. This motivates a wide range of applications from distributed filtering \cite{he2019secure}, detection \cite{aldosarimoura-jan07}, and learning \cite{TASE_quant,eaai} to networked multi-vehicle systems (NMVS) in this paper.	
Unlike centralized approaches, which are subject to single point of failure and scalability limitations, distributed algorithms facilitate real-time monitoring and decision-making across the entire NMVS. The scalability of distributed solutions is important on large-scale without significant overhead. These decentralized approaches allow for local detection and isolation of faults, minimizing the impact on overall NMVS functionality. 

Distributed or decentralized estimation and observer design refer to techniques used in control systems and signal processing where multiple interconnected agents or sensors collaboratively estimate the state of a system via partial measurements \cite{modalavalasa2021review,he2020distributed}. The scalability, robustness, and fault tolerance of these techniques make them suitable for multi-robot or multi-vehicle systems, where centralized processing is impractical or impossible. Some existing works include: 	
distributed observer under finite-time consensus \cite{fioravanti2024distributed} and event-triggered mechanism \cite{xu2022fully}, distributed estimation	
via moving horizon multi-sensor information fusion \cite{dong2022consensus,zou2020moving}, and multi time-scale distributed observers with inner consensus loop \cite{olfati:cdc09,he2020secure,he2019distributed}.	

Secure distributed estimation algorithms are crucial because distributed sensor networks are often vulnerable to cyber threats or possible faults, such as false data injection \cite{hua2022distributed,signal} and denial of service \cite{battistelli2023stability,jenabzadeh2025distributed}. Since these networks rely on cooperation and data sharing among multiple agents/nodes, a single compromised node can disrupt the entire estimation process, leading to erroneous tracking. Therefore, many works are devoted to ensuring security and reliability in these algorithms. For example,
secure diffusion-based distributed least-mean-square (LMS) estimation under multiplicative sensor attack \cite{zayyani2025secure} and adversary detection under Bayesian hypothesis test \cite{zayyani2022adversary}, graph-theoretic-based distributed observers resilient to node failure \cite{TNSE25,eurasip}, fault-tolerant event-triggered distributed observer under actuator faults \cite{hou2023fully}, and privacy-preserving distributed filtering \cite{moradi2022privacy} and subject to eavesdropping attacks \cite{yan2024privacy}.

In this work, we apply distributed estimation and observer design to NMVS and intelligent transportation systems (ITS), which enables vehicles to collaboratively estimate traffic states and vehicle positions/velocities. This collective approach improves safety and optimizes traffic flow without relying on a centralized processor and is based on distributed data fusion \cite{el2011data,ounoughi2023data}. By sharing local information such distributed algorithms enable efficient localized tracking in large-scale mixed-traffic setups. Similar algorithms with application to ITS are considered in the literature. Distributed data-driven and fault-tolerant control of connected vehicle platoons with sensor faults is discussed in \cite{zhu2023distributed}. Optimization solution applied to NMVS for traffic signal control is considered in \cite{liaquat2024assessing}. Observer-based tracking of multi-vehicle platooning is studied in \cite{jiang2021observer}. Similarly, the work \cite{zhu2024secure} studies observer-based tracking control schemes to ensure security and collision-free performance of NMVS. Moreover, fault-tolerant formation control of NMVS via finite-time convergent algorithms is discussed in \cite{wu2023finite}. What is missing in the literature is a \textit{distributed} observer-based FDI strategy for mixed-traffic tracking to enhance the safety and security of NMVS.

In this paper, we consider a mixed-traffic network of HDVs and CAVs. The main contributions are summarized as follows:
\begin{itemize}
	\item We propose a ``distributed'' observer to track the states (velocities and positions) of some HDVs via partial measurements taken by group of CAVs sharing information over a communication network. Unlike traditional centralized observers, our method enables decentralized tracking with no centralized processor, where each CAV node is able to perform computations locally. The measurement data is distributed over NMVS, and every CAV node performs \textit{local} decision-making in its neighborhood. This enhances scalability and fault-tolerance in large-scale ITS.
	\item The proposed distributed observer is of single time-scale with no inner consensus loop. This advances the existing multi time-scale methods \cite{he2019secure,olfati:cdc09,he2020secure,he2019distributed}, which require many iterations of consensus and communications between every two consecutive samples of system dynamics (referred to as the inner consensus loop). Therefore, our proposed single time-scale observer is more efficient in terms of computation/processing and communication requirements.  
	\item  We introduce an observer-based distributed fault detection strategy which is localized over the NMVS, i.e., every CAV locally finds anomaly in its observer-based residual. We make \textit{no local observability} assumption at any CAV node, but only \textit{distributed observability} over the NMVS. In ITS setup, this is important to localize the FDI and enables every CAV to locally detect possible faults at its sensor measurement with no need for central coordination, which is in contrast to centralized detection techniques.
	\item  We propose two FDI logics in this work: (i) using history of the residuals (stateful detection) and (ii) without residual history and only based on instantaneous residuals (stateless detection). For each scenario, we deisgn a \textit{probabilistic} threshold  based on noise statistics, and derive the related false-alarm-rate (FAR). We do not assume that the noise is of bounded support, which is more realistic in real-world ITS setups. This advances the detection strategies in existing literature \cite{pajic2015attack,chong2015observability,lee2015secure,kodakkadan2017observer,Riccati-weakcons} based on \textit{deterministic} threshold design. This is better explained later in Remark~\ref{rem_bound}.
\end{itemize}

\textit{Paper Organization:} Section~\ref{sec_formulat} formulates the distributed setup for mixed traffic NMVS tracking. Section~\ref{sec_fdi} presents our main results on distributed observer design and localized probabilistic-threshold-based FDI. Section~\ref{sec_sim} provides simulations to better illustrate the contributions. Section~\ref{sec_con} concludes the paper with some future research directions. 

\textit{General Notation:}
Operators ``$\circ$" and ``$\otimes$" respectively denote the entrywise (Hadamard) and Kronecker product of matrices. Matrices $I_n$ and $\mb{1}_{n\times n}$ respectively denote the identity and all-ones matrix of size $n$. $\mb{1}_{n}$ denotes the column vector of all-ones of size $n$. Operator ``$\succ$'' denotes positive-definiteness.

\section{Formulation of Multi-Vehicle Networked Tracking} \label{sec_formulat}
In this section, we consider a mixed traffic network of HDVs and CAVs and formulate a \textit{distributed} setup which CAVs track/estimate the state of all HDVs via partial sensor measurements.  Consider the following linear time-invariant (LTI) system modelling the dynamics of $N$ HDVs:
\begin{eqnarray}\label{eq_sys1} 
	\mb{x}_{k+1} = A\mb{x}_k + \nu_k,
\end{eqnarray}
with~$\mb{x}_k=(\mb{x}_{1,k};\ldots;\mb{x}_{N,k})  \in\mathbb{R}^{N m}$ (``;" denotes column concatenation) as the global state at time $k$,  $\mb{x}_{i,k} \in\mathbb{R}^{m}$ as the state of the $i$th HDV,~$A \in \mathbb{R}^{Nm\times Nm}$ as the dynamic system matrix, and~$\nu_k$ as some noise input\footnote{It is possible to also consider an input matrix $B$, i.e., random input $B\nu_k$. For example, the nearly-constant-velocity model assumes an input matrix and random unknown input variable in the form $\mb{x}_{k+1} = A\mb{x}_k + B\nu_k$}. We make standard zero-mean Gaussian assumption on the noise statistics, i.e., $\nu_k \sim \mc{N}(0,G)$. 
Each diagonal block of the system matrix $A$, denoted by $\widetilde{A}_i$ with $i\in \{1,\ldots,N\}$, represents the dynamics associated with the $i$th HDV. 

\begin{rem} \label{rem_ncv}
	The vehicle dynamics might be nonlinear, time-varying, and generally unknown. Therefore, the HDV dynamics in Eq.~\eqref{eq_sys1} is generally modelled as nearly-constant-velocity (NCV) modelling with diagonal blocks of $A$ defined as
	\begin{equation}
		\widetilde{A}_i=\left(
		\begin{array}{cc}
			1 & \mc{T} \\
			0 & 1 	
		\end{array} \right),
	\end{equation}
    or nearly-constant-acceleration (NCA) modelling with diagonal blocks of $A$ defined as
    \begin{equation}
    	\widetilde{A}_i=\left(
    	\begin{array}{ccc}
    		1 & \mc{T} & \frac{\mc{T}^2}{2} \\
    		0 & 1 	& \mc{T} \\
    		0 & 0 & 1
    	\end{array} \right),
    \end{equation}
    with $\mc{T}$ as the sampling time. Many literature on tracking and filtering use these common models for vehicle dynamics, as in \cite{zubavca2022innovative,song2017multi,blair2023mse,ennasr2020time,petersen2022tracking,tian2020ground}. This justifies the NCV model used for simulations in Section~\ref{sec_sim}, which gives acceptable performance for vehicle tracking via distributed observer design. However, on the other hand, if the exact vehicle dynamics is known, one can linearize this model and apply the linearized $A$ matrix in Eq.~\eqref{eq_sys1}. 
\end{rem}

Each of $n$ CAVs is embedded with certain sensors to take state measurements of few HDVs.\footnote{It is possible to extend to the case that few number of CAVs have no sensor measurements of any HDV. In this case, these CAVs can track the states of HDVs only based on the information shared by their neighboring CAVs. } The sensor measurements are in the following form
\begin{align} \label{eq_H_i}
	\mb{y}_{i,k} = C_i\mb{x}_k + {\mu}_{i,k},~~~i=1,\dots,n
\end{align}
with $\mb{y}_{i,k}\in\mathbb{R}^{l_i}, ~C_i\in\mathbb{R}^{l_i\times Nm}$, and~${\mu}_{i,k}$  as the measurement vector, local output matrix, and noise at the $i$th CAV, respectively.  We make standard assumptions on Gaussianity and statistical independence of the noise terms, i.e., $\mathbb{E}({\mu}_{i,k}) = 0$ and $\mathbb{E}( {\mu}_{i,k} {\mu}_{i,m}) = 0$.  
The overall measurement vector (or global output vector) of the entire group of CAVs is 
\begin{align} \label{eq_H}
	\mb{y}_k = C\mb{x}_k + \mu_k,
\end{align}
with~$\mb{y}_k \in\mathbb{R}^{L},~C=(C_1;C_2;\ldots;C_n)$  as \textit{global} output matrix
with~$L=l_1+\ldots+l_n$ and~$\mu_k \sim \mc{N}(0,R)$ as the global noise. Without loss of generality, we assume single state measurements, i.e., $l_1=\ldots=l_n=1$ and $L=n$, however, the results of this paper can be easily extended to the more general case. To satisfy the necessary condition for observability, it is assumed that the state of every HDV is measured by (at least) one CAV; otherwise, the observability condition is violated and some HDVs cannot be tracked.  

There are two main scenarios for estimating or tracking the state of HDVs: (i) \textit{Centralized filtering} in which all sensor measurements are provided to a central unit/processor to filter the noisy sensor outputs, given that the pair $(A,C)$ is observable \cite{bay}. Examples of such centralized observer design are given in the literature \cite{biroon2021false,zhao2024observer,rostami2020state,liu2024modified}. (ii) \textit{Distributed filtering} where each CAV has a local filtering unit such that, by communicating over an information-sharing network $\mc{G}$, each CAV can estimate the global state, ${\mb{x}}_{k}$, given the local sensor measurements and a-priori estimates from other vehicles in its neighborhood $\mc{N}(i)$. In this scenario, the challenge is that the entire HDV system is not observable to any single CAV (no local $(A,C_i)$-observability) and \textit{distributed observability} over the entire NMVS is assumed. By satisfying distributed observability, the distributed/networked observer enables each CAV to track the HDVs only by local data exchange over the network. Some examples of existing distributed observer/estimator for LTI systems are given in \cite{modalavalasa2021review,TNSE25,abdelmawgoud2020distributed,fioravanti2024distributed,he2020distributed,mitra2018distributed}. In this paper, the condition for distributed observability is established as follows, which is much less restrictive than the local observability assumption.
\begin{defn} \label{defn_1}
	Given a network $\mc{G}$ of $n$ CAVs estimating the state of $N$ HDVs with dynamics~\eqref{eq_sys1} and outputs~\eqref{eq_H} the \textit{distributed observability} is defined as observability of the following pair
	\begin{equation} \label{eq_dist_obsrv}
		(W \otimes A, D_C),
	\end{equation}
	where $W \in \mathbb{R}^{n\times n}$ denotes the consensus adjacency matrix following the structure of the CAVs' communication network $\mc{G}$, $A$ is the system matrix of HDVs representing their dynamics, and $D_C$ represents the shared observation matrix defined as
	\begin{eqnarray} \label{D_H}
		D_C =
		\left(
		\begin{array}{ccc}
			\sum_{j\in\mc{N}(1)}C_j^\top C_j&&\\
			&\ddots&\\
			&&\sum_{j\in\mc{N}(n)}C_j^\top C_j
		\end{array}
		\right).
	\end{eqnarray}		 
\end{defn}

\begin{figure} 
	\centering
	\includegraphics[width=4in]{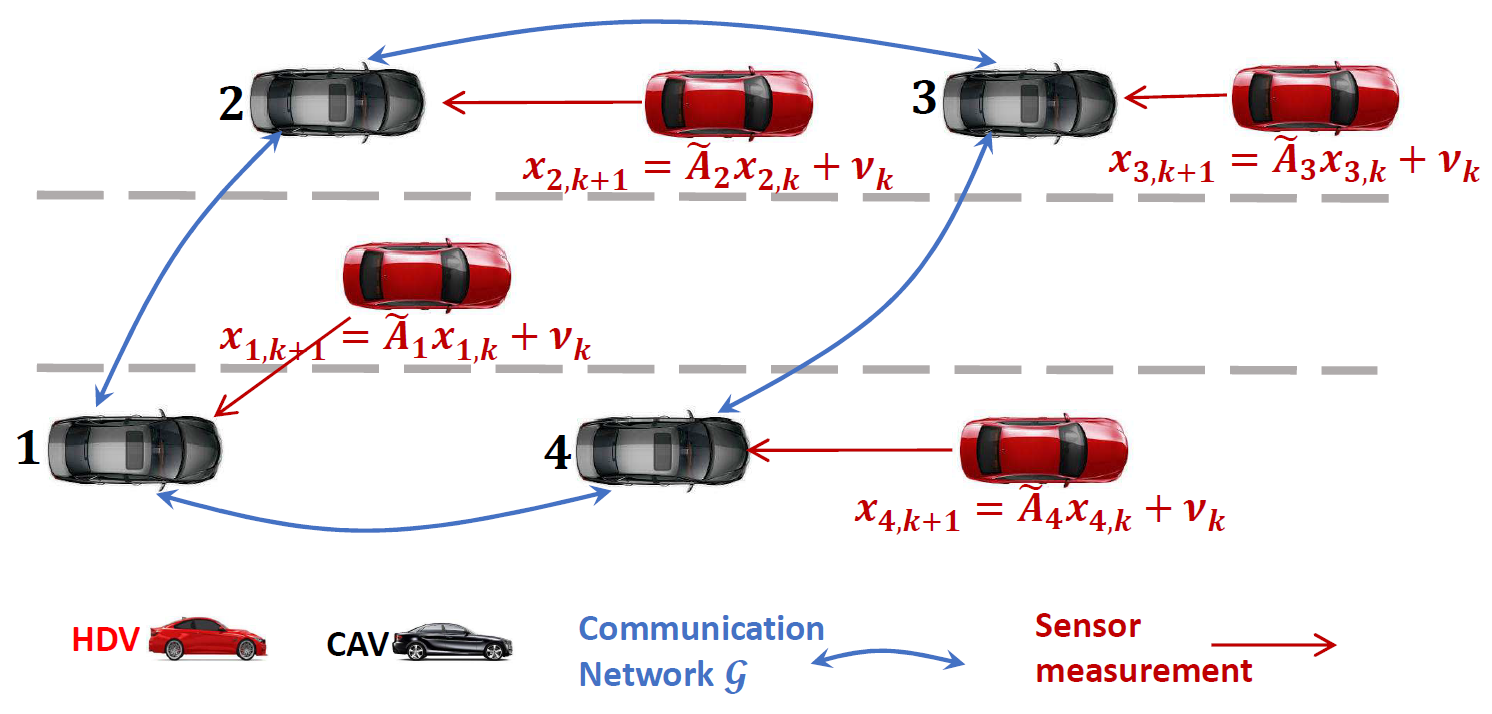}
	\caption{This figure shows a mixed traffic NMVS of 4 HDVs and 4 CAVs. The idea is to enable every CAV to track the state of the HDVs via a distributed observer. In this example, each CAV only has partial sensor measurement of 1 HDV and not the entire group of HDVs. The distributed observer enables every CAV to estimate the entire state of \textit{all} HDVs, denoted by $\mb{x}_k$ in this paper.
	} \label{fig_platoon}
\end{figure}

A better illustration of this distributed problem setup is given in Fig.~\ref{fig_platoon}.
The main concern is that the measurement data at some CAVs might be subject to faults, biases, anomalies, or attacks and are not trustworthy. Therefore, the faulty CAV sensors need to be locally detected and isolated to avoid the cascade of faults over the entire NMVS. However, it is assumed that (at least) one trustworthy sensor output from each HDV is available to the group of CAVs; otherwise, no observer design results in fault-free tracking. In this work, the idea is to perform distributed observer-based FDI to locally detect and isolate the faulty CAV sensor. For this, first, the distributed estimator/observer is given and, then, localized FDI algorithms are proposed to track the so-called \textit{residual} and locally detect possible faults at every CAV. 

\section{Distributed and Localized Fault-Detection Algorithms} \label{sec_fdi}
\subsection{The Proposed Distributed Observer} \label{sec_observer}
In this section, we first propose our distributed observer consisting of two steps: (i) \textit{Predicton step} with consensus on a-priori estimates and (ii) \textit{Innovation update} with sharing measurements. The following formulation describes the local observer at vehicle $i$:
\begin{eqnarray}\label{eq_p}
	\widehat{\mb{x}}^i_{k|k-1} &=& \sum_{j\in\mathcal{N}(i)} W_{ij}A\widehat{\mb{x}}^j_{k-1|k-1},
	\\\label{eq_m}
	\widehat{\mb{x}}^i_{k|k} &=& \widehat{\mb{x}}^i_{k|k-1} + K_i \sum_{j\in \mc{N}(i)}C_j^\top \left(\mb{y}_{j,k}-C_j\widehat{\mb{x}}^i_{k|k-1}\right),
\end{eqnarray}
where $\widehat{\mb{x}}^i_{k|k}$ denotes the estimate of $\mb{x}_k$ as the HDVs' states given all the sensor measurements up to time $k$. Define $\mathcal{N}(i)$ as the neighborhood of CAV $i$ over which the a-priori estimates and measurements are shared. The neighborhood includes the vehicle $i$ itself implying that the vehicle uses its own estimation and measurement data. The matrix $K_i$ denotes the gain at vehicle $i$ and needs to be designed properly to stabilize the observer error. The matrix $W$ denotes the consensus matrix for averaging a-priori estimates with $W_{ij}$ as the consensus weight at the data of CAV $j$ delivered to CAV $i$. This matrix $W$ is row-stochastic to satisfy the consensus requirement, i.e.,
\begin{align}
	\sum_{j=1}^n W_{ij} = 1,~\mbox{or}~\sum_{j \in \mc{N}(i)} W_{ij} = 1.
\end{align} 

Other than row-stochasticity, no other assumption is needed on the consensus weights and, therefore, the weights can be chosen simply as $\frac{1}{|\mc{N}(i)|}$, random stochastic, or based on, for example, the Metropolis-Hastings fusion rule \cite{Xiao05distributedaverage}. The structure of matrix $W$ follows the structure/topology of the vehicular network $\mc{G}$. 

\begin{rem}
	The proposed distributed observer \eqref{eq_p}-\eqref{eq_m} outperforms the existing multi time-scale distributed observers in \cite{olfati:cdc09,he2020secure,he2019distributed,battistelli2023stability,dong2022consensus,zou2020moving} with inner consensus loop. This is because, in these works, every node needs to perform $L$ steps of consensus data fusion and communications between every two consecutive time iterations $k$ and $k+1$ of dynamics~\eqref{eq_sys1}. In terms of practical application, this implies $L$ times faster communication and processing units at vehicles. However, our single time-scale distributed observer only requires $1$ step of consensus/communication between two consecutive time iterations $k$ and $k+1$. This significantly reduces the communication and processing load on the vehicular network. 
\end{rem}

Given that $(A,C)$ observability holds, there exists cone-complementary LMI gain design for $K=\mbox{blockdiag}[K_i]$ to ensure $(W \otimes A, D_C)$ (as described in Definition~\ref{defn_1}). These cone-complementary LMI algorithms are of polynomial-order complexity and are computationally efficient. Some of these algorithms can be found in \cite{rami:97,usman_cdc:11}. In this paper, an example is given later in Algorithm~\ref{alg_lmi}. 

Define the observer error at CAV $i$ as ${\mb{e}_{k}^i := \mb{x}_{k} - \widehat{\mb{x}}^i_{k|k}}$ and the gloabl error vector at all CAVs as ${\mb{e}_{k} = (\mb{e}_{k}^{1}; \dots;\mb{e}_{k}^N)}$ with ``;" as column concatenation. The global error dynamics for the observer \eqref{eq_p}-\eqref{eq_m} is 
\begin{align} \nonumber
	\mb{e}_{k} &= (W\otimes A - KD_C(W\otimes A))\mb{e}_{k-1} +
	{\eta}_k \\  
	&= \widehat{A}\mb{e}_{k-1} +
	{\eta}_k, \label{eq_err1}
\end{align}
with $\widehat{A} := W\otimes A - KD_C(W\otimes A)$ as the closed-loop system matrix and ${\eta}_k$ collecting the noise terms as,
\begin{align}
	{\eta}_k &= \mathbf{1}_n \otimes {\nu}_{k-1} 
	- K D_C(\mathbf{1}_n \otimes {\nu}_{k-1}) - K\overline{D}_C{\mu}_{k}, 
	\label{eq_eta}
\end{align}
where  
\begin{eqnarray}\nonumber
	\overline{D}_C &:=& \left(
	\begin{array}{ccc}
		W_{11} C_1^\top&  \dots & W_{1n} C_n^\top\\
		&\ddots\\
		W_{n1} C_1^\top&  \dots & W_{nn} C_n^\top
	\end{array}
	\right) \\
	&:=& (W \otimes \mb{1}_N) \circ (\mb{1}_n \otimes C^\top).
\end{eqnarray}
Given the error dynamics~\eqref{eq_err1}, the necessary condition for steady-state error stability is the observability of the pair ${(W\otimes A,D_C)}$, which is called \textit{distributed observability} in Definition~\ref{defn_1}. This follows the well-known Kalman stability theorem. We refer interested readers to \cite{bay,usman_cdc:11} for more discussion on the error stability of linear observers. 

Next, we need to properly design the structure of the CAV network $\mc{G}$ (i.e., the structure of the consensus matrix $W$) to ensure ${(W\otimes A,D_C)}$-observability. This is done by structural system analysis \cite{ramos2022overview}. Let denote the structure of the system matrix $A$ by the system graph $\mc{G}_A$. Then, the observability of ${(W\otimes A,D_C)}$ follows the structural observability of \textit{Kronecker product} of the networks $\mc{G}$ and $\mc{G}_A$. Using the results from Kronecker composite network observability defined in \cite[Theorem~4]{kronecker_TSIPN}, the sufficient observability condition is that the network $\mathcal{G}$ be strongly-connected\footnote{The minimal network satisfying strong-connectivity is a cycle network; one such example sparse network is shown in Fig.~\ref{fig_platoon}. In general, however, more dense networks can be adopted to have more information-sharing between CAVs to design more resilient distributed estimators. This follows the context of $q$-node/link-connectivity and implies that the distributed estimator is resilient to removal of up-to $q$ communication-links/processing-nodes.}. We refer interested readers to \cite{kronecker_TSIPN,cartesian_TSIPN} for more information. Table~\ref{tab_compare} compares the observability assumption and network connectivity requirement in this paper with some state-of-the-art literature on distributed tracking.
\begin{table} [h] 
	\centering
	\caption{Comparing the observability assumption and network connectivity requirement }
	\label{tab_compare}
	\scalebox{0.8}{\begin{tabular}{|c|c|c|c|} 
		\hline
		Ref. & Observability  
		& Formulation
		 & Minimum links\\   
		\hline
		\cite{sauter:09} &  global &
		$(A,C)$-observability &  $n(n-1)$ 
		\\
		\hline
		\cite{he2019distributed,hua2022distributed,zayyani2025secure,zayyani2022adversary,abdelmawgoud2020distributed,mitra2018distributed} &  local &
		$(A,C_i)$-observability &  $N(n-1)$,~$N \leq n$ 
		\\
		\hline
		This work &  distributed &
		$(W\otimes A,D_C)$-observability &  $n-1$ 
		\\
		\hline
		\hline
	\end{tabular}}
\end{table}

Given that  $\mathcal{G}$ is strongly-connected, one can find proper feedback gain $K$ such that ${\rho(\widehat{A})<1}$ (with $\rho(\cdot)$ denoting the spectral radius); this ensures steady-state stability of the error dynamics~\eqref{eq_err1}. Note that the network of CAVs can be switching over time as long as it is strongly-connected to satisfy the sufficient condition for distributed observability. However, in practice, the switching period should be large enough to allow the transient error to settle and reach steady-state performance.

\begin{rem} \label{rem_comunic}
 The communication complexity of the proposed observer design is $\mc{O}(n)$ with $n$ as the number of CAVs. This is because a strongly-connected network $\mc{G}$ (with minimum number of $n$ links) suffices for the distributed observability and stability of the distributed observer. This implies that the distributed observer design can be scaled up for large-scale setups. 
\end{rem}

\subsection{Distributed Fault Detection via Instantaneous Residuals}
In this section, we assume that certain CAVs may encounter faults (anomalies or even attacks) denoted by $f_{i,k}$ in their sensor measurements. Then, the sensor measurement at CAV $i$ at time $k$ is	 
\begin{align} \label{eq_f_i}
	\mb{y}_{i,k} = C_i\mb{x}_k + {\mu}_{i,k}+f_{i,k},~~~i=1,\dots,n
\end{align}
The idea is to locally detect and isolate these faulty measurements with no need for a centralized coordinator/unit. For this purpose, we track the output residual at every CAV. Given the state estimation $\widehat{\mb{x}}_{k|k}^i$ at CAV $i$, the estimated measurement is defined as $\widehat{\mb{y}}_{i,k}= C_i\widehat{\mb{x}}_{k|k}^i$. Then, the residual is defined as,
\begin{eqnarray}\nonumber
	\mb{r}_k^i &=& |\mb{y}_{i,k}-\widehat{\mb{y}}_{i,k}|=|\mb{y}_{i,k}-C_i \widehat{\mb{x}}_{k|k}^i| \\ \nonumber &=& |C_i \mb{e}_k^i+{\mu}_{i,k}+f_{i,k}|\\
	&=& |C_i \widehat{A}_i \mb{e}^i_{k-1}+C_i {\eta}_{i,k}+{\mu}_{i,k}+f_{i,k}|,
	\label{eq_r}
\end{eqnarray}
with $\widehat{A}_i$ as the block of rows $(i-1)n+1$ to $in$ of $\widehat{A}$ (the $i$th hyper-row of $\widehat{A}$). By tracking this instantaneous residual, one can detect possible faults at CAV sensors. In this subsection,  detection decisions are made instantaneously by examining the current residual only, without considering the past history of residual values. This is also referred to as the \textit{stateless} detection \cite{giraldo2018survey}. In terms of implementation, this method is simple and with no need for memory or storage of past residuals. On the other hand, instantaneous residuals can be noisy, causing false alarms or missed detections if residual spikes are transient. This is also more sensitive to fluctuations and single outliers.

When $f_{i,k}=0$ for all $i$ (i.e., the fault-free case), the observer error $\mb{e}_k^i$ and its associated residual $\mb{r}_k^i$ are bounded and steady-state stable for all CAVs. This is because $\widehat{A}$ is Schur stable as illustrated in Section~\ref{sec_observer} and, in steady-state, the first term in \eqref{eq_r} goes to zero, while the remaining terms affecting the residual are $C_i{\eta}_{i,k} + {\mu}_{i,k}+f_{i,k}$. We have,
\begin{align} \nonumber
	C_i{\eta}_{i,k} &= C_i\mb{\nu}_{k-1} \\ \label{eq_r_eta} &-C_iK_i\sum_{j\in \mathcal{N}(i)} \Bigl(C_j^\top {\mu}_{j,k} +
	C_j^\top {f}_{j,k}+ C_j^\top C_j\mb{\nu}_{k-1} \Bigr).
\end{align}

In faulty case where ${f}_{j,k} \neq 0$ at CAV $j$ we have $C_i K_i C_j^\top {f}_{j,k} \neq 0$ at every CAV $i \in \mc{N}(j)$, i.e., at all neighboring CAVs $j$. Therefore, the fault not only affects the residual at the same CAV sensor but also changes the residual at all the neighboring sensors. This implies that if the residual at some CAVs is biased more than certain predefined \textit{thresholds}, one or more CAV sensors are faulty.
Next, we aim to isolate the faults to specifically determine which CAV sensor is faulty. For this purpose,  one needs to design the gain matrix $K$ such that the residual at the faulty sensor is biased considerably more than other neighboring CAVs.

\subsubsection{Gain Design} 
Given $ \mc{K}=I_n \otimes \mb{1}_{N\times N}$ and sufficiently small parameter $\epsilon$ to scale the residuals,
one can design block-diagonal gain matrix $K=K \circ \mc{K}=\mbox{diag}[K_i]$ from Algorithm~\ref{alg_lmi}.
For isolation of the faulty CAV sensor, the gain matrix $K$ in Algorithm~\ref{alg_lmi} satisfies,
\begin{align} \label{eq_Kalpha}
	\left |\frac{C^\top_i K_i C_j}{C^\top_j K_j C_j-1} \right | \leq \epsilon, ~ \mbox{for}~ i \neq j , \forall j \in \mc{N}(i)
\end{align}
with $0 < {\epsilon < 1}$ as a predefined scale parameter determining the residual ratio between the faulty CAV $j$ and its neighbors $i$.
Assuming non-zero fault $f_{k,j}\neq 0$, Eq. \eqref{eq_r}-\eqref{eq_r_eta} implies that the residual $\mb{r}_k^i$ at CAV $i$ with ${i \in \mathcal{N}(j)}$ is affected via the term $C^\top_i K_i C_j$, while the residual $\mb{r}_k^j$ at CAV $j$ 
is affected via $C^\top_j K_j C_j-1$. 
Then, for ${i \neq j}$, Eq. \eqref{eq_Kalpha} guarantees that $${\frac{\mb{r}_k^j}{\mb{r}_k^i}>\frac{1}{\epsilon} > 1}.$$ Therefore, the residual at CAV $j$ is greater by a factor $\frac{1}{\epsilon}$ than the residual at CAV $i$, which implies that one can isolate the fault/bias at CAV $j$ for sufficiently small $\epsilon$.

\begin{rem} \label{rem_comp}
	The LMI-based Algorithm~\ref{alg_lmi} can be solved via iterative cone-complementary optimizations, e.g., see \cite{rami:97, siljak08} for details. It is known that these cone-complementary optimization algorithms are of polynomial-order complexity  \cite{ye1993fully,nesterov1994interior}. Furthermore, the complexity of the proposed observer and residual evaluations is of the order of $\mc{O}(n^2N^2m^2)$. The polynomial-order complexity implies that the distributed solution can be scaled up for large-scale setups. 
\end{rem}

\begin{algorithm} [t] \label{alg_lmi}
	\textbf{Input:} matrices $A$, $W$, $D_C$, scale parameter $\epsilon$ \\ 
	Iteratively solve for $\widehat{A} = W\otimes A - KD_C(W\otimes A)$:  
	\begin{align}
		\begin{aligned}
			\displaystyle
			\min
			~~ &  \mathbf{trace}(XY) \\
			\text{s.t.}  ~~& X,Y\succ 0,  & K \leftarrow  K \circ \mc{K} 
			\\ \nonumber ~ & \left( \begin{array}{cc} X&\widehat{A}^\top\\ \widehat{A}&Y\\ \end{array} \right) \succ 0,~& \left( \begin{array}{cc} X&I\\ I&Y\\ \end{array} \right) \succ 0,
		\end{aligned}\\ \nonumber
		\frac{|C_i K_i C_j^\top|}{|1-C_j K_j C_j^\top |}<\epsilon,~\forall j \in \mc{N}(i),~j \neq i
	\end{align}
	
	\textbf{Output} Block-diagonal gain $K$\;	
	\caption{Constrained LMI gain design} 
\end{algorithm}

\subsubsection{Probabilistic Threshold Design}
In this subsection, we define the thresholds based on which additive biasing faults are detected.
Let $P_k := \mathbb{E}(\mb{e}_k\mb{e}_k^{\top})$ and $Q :=\mathbb{E}(\eta_k\eta^\top_k) $. From error dynamics \eqref{eq_err1},
\begin{eqnarray} \nonumber
	P_{k} &=& \widehat{A}P_{k-1}\widehat{A}^\top + Q \\
	&=& \widehat{A}^kP_{0}(\widehat{A}^\top)^k + \sum_{j=0}^{k-1}  \widehat{A}^{j}Q(\widehat{A}^\top)^{j}.
\end{eqnarray}
Recall that $\widehat{A}$ is Schur stable based on the distributed observer design, and in steady-state we have,
\begin{eqnarray} \nonumber
	\lim_{k \rightarrow \infty} \widehat{A}^kP_{0}(\widehat{A}^\top)^k \rightarrow 0,
\end{eqnarray}
Define $P_{\infty} := \lim_{k \rightarrow \infty} P_{k}$. One can conclude that,
\begin{eqnarray} \nonumber
	P_{\infty} =  \sum_{j=0}^{\infty}  \widehat{A}^{j}Q(\widehat{A}^\top)^{j}.
\end{eqnarray}
Let $\|\widehat{A}\|_2=\beta $ where we have $\beta<1$. Then, borrowing some results from \cite{khan2014collaborative}, we have
\begin{eqnarray} \label{eq_pinfty1}
	\|P_{\infty}\|_2  \leq \sum_{j=0}^{\infty} \| \widehat{A}^{j}Q(\widehat{A}^\top)^{j} \| \leq \sum_{j=0}^{\infty} \beta^{2j}\|Q\|_2 \leq \frac{\|Q\|_2}{1-\beta^2},
\end{eqnarray}
where $\|P_{\infty}\|_2$ is the $2$-norm of the covariance of the error $\mb{e}_k$ as $k \rightarrow \infty$.
In the absence of faults, i.e., $f_{i,k} = 0$ for all $i$, we have
\begin{align} \nonumber
	\eta_k\eta^\top_k =& (I_{Nn}- K D_C)(\mathbf{1}_{n\times n} \otimes \nu_{k-1}\nu_{k-1}^\top)(I_{Nn}- K D_C)^\top\\ &+ (K\overline{D}_C) \mu_k\mu_k^\top (K\overline{D}_C)^\top.
\end{align}
Let $\mathbb{E}(\nu_{k-1}\nu_{k-1}^\top)=G$ and $\mathbb{E}(\mu_k\mu_k^\top)=R$. Using the {$\mathbb{E}$-operator} on both sides of the above equation we get,
\begin{eqnarray} \nonumber
	Q &=& (I_{Nn}- K D_C)(\mathbf{1}_{n\times n} \otimes G)(I_{Nn}- K D_C)^\top\\ &+& (K\overline{D}_C) R (K\overline{D}_C)^\top.
\end{eqnarray}
Applying the $2$-norm operation,
\begin{eqnarray} \nonumber
	\|Q\|_2 &\leq& \|(I_{Nn}- K D_C)(\mathbf{1}_{n \times n} \otimes G)(I_{Nn}- K D_C)^\top\|_2\\ \nonumber
	&+& \|(K\overline{D}_C) R (K\overline{D}_C)^\top\|_2 \\
	&\leq& \|I_{Nn}- K D_C\|_2^2 N \|G\|_2 + \|K\|_2^2 \|\overline{R}\|_2,
\end{eqnarray}
with
\begin{eqnarray}\nonumber
	\overline{R} := \left(
	\begin{array}{ccc}
		\sum_{i\in \mathcal{N}(1)} C_i^\top R_i C_i\\
		&\ddots\\
		& &\sum_{i\in \mathcal{N}(n)} C_i^\top R_i C_i\
	\end{array}
	\right).
\end{eqnarray}
Let $\alpha_1 := \|I_{Nn}- K D_C\|_2^2$ and $\alpha_2 := \|K\|_2^2$. Then, recalling \eqref{eq_pinfty1}, we have
\begin{eqnarray} \label{eq_pinfty}
	\frac{\|P_{\infty}\|_2}{n}  \leq \frac{\alpha_1n\|G\|_2+\alpha_2 \|\overline{R}\|_2}{n(1-\beta^2)}=\Theta.
\end{eqnarray}
Note that, in the above equation, $\|P_{\infty}\|_2$ is rescaled by the total number of CAVs $n$ to find the $2$-norm of the covariance of the local error $\mb{e}_k^i$ at every CAV. Eq.~\eqref{eq_pinfty} in fact gives the upper-bound $\Theta$ on the variance of $\mb{e}_k^i$ at CAV $i$. 

\begin{figure} 
	\centering
	\includegraphics[width=3.5in]{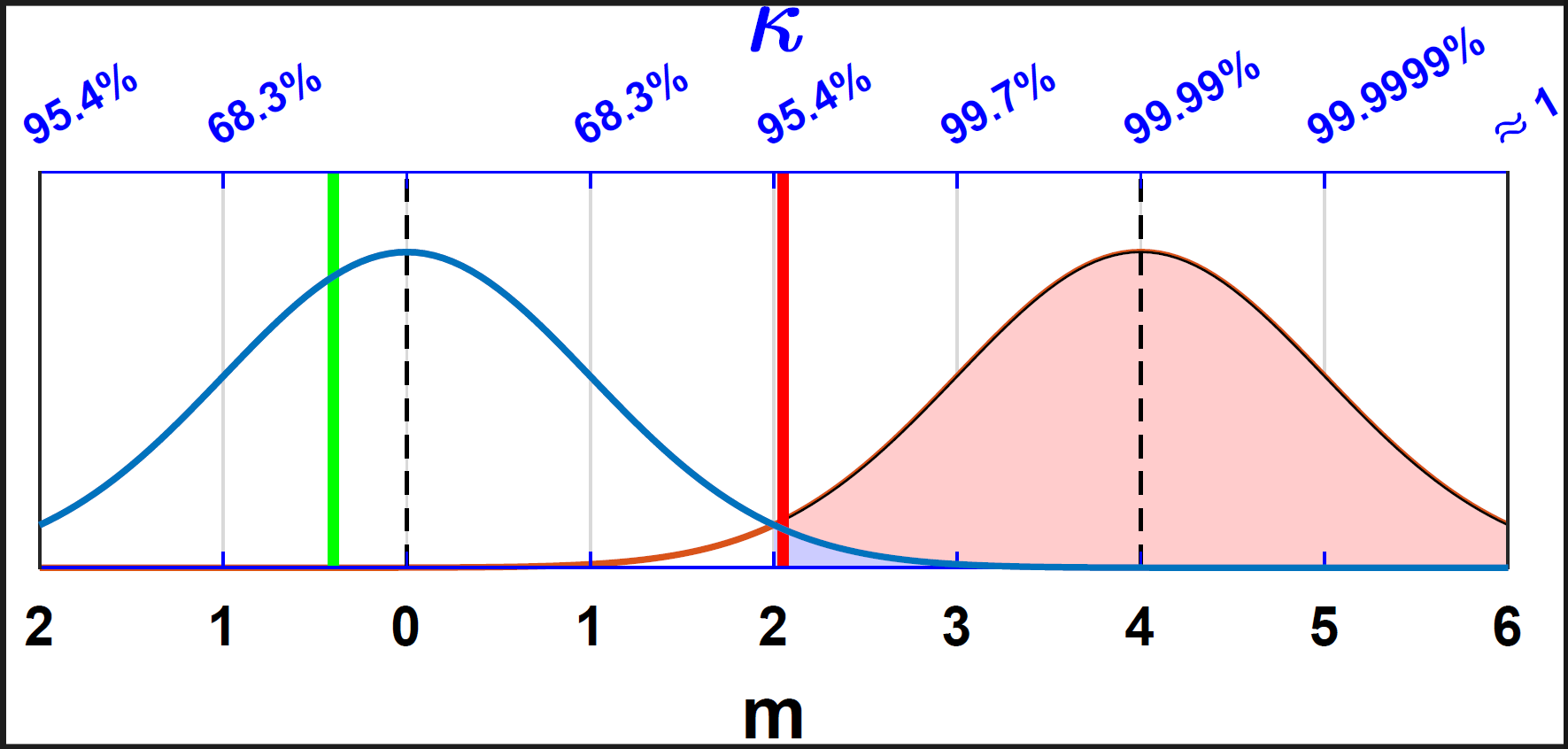}
	\caption{This figure illustrates the FDI strategy in this paper. For the normalized residual in the fault-free case (blue curve), the confidence intervals are given. The $m$ value in Eq. \eqref{eq_thresold} denotes a confidence interval and is associated with a probabilistic threshold $\kappa$.  Two example red and green lines are considered as two normalized residual values $\frac{r_k}{\Phi}$. From the binary hypothesis testing, the threshold is the intersection of the two PDFs, the fault-free case (blue curve) and the faulty case (red curve). From the figure, the red line exceeds the threshold $\theta_\kappa$ with $m=2$ and, thus, the fault probability is more than ${\kappa = 95.4\%}$. This probability equals the red shaded area and the false alarm rate is $1-\kappa=4.6\%$, which is the area shaded by blue. The green residual with ${r_k<{\Phi}}$ is likely due to system/output noise, as it is evident from the blue PDF curve. 
	} \label{fig_2normal}
\end{figure}

Recall that in the absence of faults, we have $\mathbb{E}(\mb{e}_k^i )=0$ in steady state. Then, following the standard Gaussian assumption and the fact that $\Theta$ is the bound on the error variance, one can design the probabilistic thresholds. Having $c$ as the measurement gain, the FDI threshold for a  detection-level ${m \in \mathbb{R}_{>0}}$ is\footnote{The probabilistic threshold design depends on the knowledge of system and measurement noise covariances. In many existing literature, it is assumed that noise covariances are known or can be approximated, based on which the thresholds are designed; see \cite{liu2017sensor,deng2025fault,samet2021statistical,mwongera2023fault,chatterjee2017improved} for example. However, covariance mismatch may affect the threshold design and FAR. For this, one can consider underestimated and overestimated noise covariances to provide both tight and conservative FAR approximations. },
\begin{align} \label{eq_thresold}
	\theta_\kappa \coloneqq m\Phi,~\Phi \coloneqq c\Theta + R.
\end{align}
with ${\kappa = \mbox{erf}(\frac{m}{\sqrt{2}})}$
as the detection probability and $\mbox{erf}(\cdot)$ as the \textit{Gauss error function}. This follows from standard  Gaussian distribution results, e.g., see \cite{krishnamoorthy2016handbook}, and is better illustrated in Fig.~\ref{fig_2normal}.
Some typical values for parameter $m$ in \eqref{eq_thresold} and the associated fault probability are given below.
\begin{itemize}
	\item $\mb{r}_k^i>\Phi$ implies fault probability more than $68\%$;
	\item  $\mb{r}_k^i>2\Phi$ implies fault probability more than $95\%$;
	\item  $\mb{r}_k^i>3\Phi$ implies fault probability more than $99.7\%$;
\end{itemize}
Note that one can design the threshold $\theta_\kappa$ based on FAR $\varkappa = 1-\kappa$ as ${\theta_\kappa = \sqrt{2}\mbox{erf}^{-1}(\kappa)\Phi}$.
Larger $f_{i,k}$ values result in larger residual $\mb{r}_k^i$ and, in turn, exceeding higher probability threshold $\theta_\kappa$ with lower FAR ${\varkappa = 1-\kappa}$.

\begin{rem} \label{rem_bound}
	Two main assumptions on the noise model exist in the literature. While some works consider Gaussian noise models as in this paper, some existing fault/attack detection literature assume that the noise term is of bounded-support, i.e., $\| \mu_{i,k}\|<\overline{\mu}$ and $\| \nu_k\|<\overline{\nu}$, with $\overline{\mu}$ and $\overline{\nu}$ as some upper-bounds, see for example \cite{pajic2015attack,chong2015observability,lee2015secure,kodakkadan2017observer,Riccati-weakcons}. In this paper, we make no such simplified assumption and consider the most general and standard form for the noise terms.
\end{rem}

	\begin{rem} \label{rem_adaptive}
		In this paper, we adopted a probabilistic threshold design for the fault detection process, where the results can be extended to use \textit{adaptive} threshold design. Adaptive thresholding adjusts detection thresholds dynamically based on statistical information, where the Gaussian noise covariance may change over time. This methodology may reduce FAR by accounting for local variations in noise, uncertainties, and system operating point. For example, the work \cite{raka2013fault} considers a dynamic interval approach for the fast computation of robust adaptive thresholds for uncertain linear systems. Adaptive energy thresholds for fault detection in active distribution networks are introduced in \cite{santos2022efficient}, aiming to achieve higher sensitivity. A self-adaptive threshold based on Hilbert-Huang transform and sliding window strategy is provided in \cite{li2021hilbert} to increase the detection sensitivity over microgrids. Residual-based robust fault detection using adaptive threhsold considering parametric modelling uncertainty using interval models is discussed in \cite{puig2013adaptive}. Time-varying adaptive threshold-based fault detection for SISO systems in strict-feedback form with unmeasurable states and
		uncertainties is presented in \cite{zhang2016time}.
    \end{rem}

\subsubsection{Main FDI Algorithm}
Following the FDI threshold design in the previous subsection, the residual $\mb{r}_k^i$ at the faulty CAV $i$ is more biased over certain probabilistic thresholds $\theta_\kappa$ (given by Eq. \eqref{eq_thresold}) than the residuals at other fault-free CAVs. The largest value of $\kappa$ satisfying $\mb{r}_k^i \geq \theta_\kappa$ gives the fault probability, and the associated FAR is ${1-\kappa}$. The FDI logic might be designed based on \textit{given} FAR ${\varkappa_i}$  at  CAV $i$. Then, for this FAR ${\varkappa_i}$, the following hypothesis testing locally declares ``Fault Alarm" or ``No-Fault" at that CAV $i$ with probability $\kappa$,
\begin{equation}
	\text{If}~ \left\{
	\begin{array}{@{}l}
		\mb{r}_{k}^i\geq \theta_{\kappa_i} \\
		\mb{r}_{k}^i<\theta_{\kappa_i} 
	\end{array}\right. ~\text{Then}~\left\{
	\begin{array}{@{}l}
		\mc{H}^i_1: \text{Fault Alarm} \\
		\mc{H}^i_0: \text{No Fault} 
	\end{array}\right. 
\end{equation}

The proposed local FDI strategy is summarized in Algorithm~\ref{alg_logic}.
\begin{algorithm} [t] \label{alg_logic}
	\textbf{Input:} $A$, $W$, $C$, $\mb{y}$, FAR $\varkappa$ (or detection probability $\kappa$), $K$ from Algorithm~\ref{alg_lmi} \\
	Find $\mb{r}_k^i$ at CAV $i$ via \eqref{eq_r}\;
	Find $\Phi$ and $\Theta$ from Eq.~\eqref{eq_thresold} and \eqref{eq_pinfty}\;
	Define probabilistic threshold $\theta_\kappa = m\Phi $\;
	\Begin(Stateless detection at CAV $i$:){
		\If{ $\mb{r}_k^i\geq \theta_\kappa$}
		{Declare: fault alarm\;}}
	\textbf{Output} Binary decision (Alarm or no-Alarm) \;
	\caption{Stateless Residual-based FDI of NMVS}
\end{algorithm}
This algorithm locally finds the faulty sensor measurement (or the associated CAV) over the NMVS. This faulty node is isolated from the rest of the CAV network $\mc{G}$. One can predesign the network $\mc{G}$ such that it is resilient to such node removal. This is by adding link redundancy in the network and is referred to as \textit{survivable network design} \cite{kerivin2005design}. This is done via graph-theoretic algorithms based on the notion of $q$-node-connectivity\footnote{A network/graph is $q$-node connected if it remains strongly-connected after removal of \textit{any} set of up-to $q$ nodes. This holds for any arbitrary set of up-to $q$ nodes with no constraint on the node selection.}. By such a resilient network design, the distributed observer tracks the state of the HDVs over the remaining network after the removal/isolation of the faulty sensor. In this paper, we assume no packet drop; however, $q$-link-connected design of the network can improve the resiliency of the distributed observer to link removal or packet drops. By adding redundant links (and more information sharing)  the network preserves strong-connectivity even after losing up-to $q$ links (possibly due to packet drops); this implies that the distributed observability holds and tracking capability can be preserved. This is a direction of our future research.

\subsection{Distributed Fault Detection via Residual History} 
In this subsection, an FDI strategy is given considering the history of the residuals $\mb{r}_k^i$ over a \textit{sliding time-window} $T$.	This is also referred to as the \textit{stateful} detection \cite{giraldo2018survey}. This method monitors past residuals and integrates temporal information to make FDI decisions, which reduces the effect of transient noise spikes. This scenario can detect subtle faults that cause consistent, moderate residual increases rather than single large spikes.
However, this comes with increased complexity and requires memory to store past residuals. Moreover, fault detection is delayed by the window length and FDI speed decreases with larger $T$. This implies a trade-off between window length and FDI delay to balance sensitivity and false alarms.

For this case, define the \textit{distance measure} as,
\begin{equation} \label{eq_z}
	\psi_{i,k}^T = \sum_{m=k-T+1}^k \frac{(\mb{r}_m^i)^2}{\Phi}.
\end{equation}
This is defined over a {sliding time-window} of length $T$. From statistics theory, the associated distribution of the summation of squared random normal variables (i.e., $\frac{(\mb{r}_m^i)^2}{\Phi}$) follows the so-called \textit{Chi-square} distribution with $T$ degrees of freedom (denoted by $\chi^2_T$) \cite{giraldo2018survey}, where $\mathbb{E}(\psi_i^T)=T$. We design the threshold similar to the stateless detection. For a prespecified FAR $\varkappa$ the probabilistic threshold is
\begin{align} \label{eq_theta_T}
	\theta_\varkappa^T = 2\Gamma^{-1}(1-\varkappa,\frac{T}{2}),
\end{align}
with $\Gamma^{-1}(\cdot,\cdot)$ as the \textit{inverse regularized lower incomplete gamma function}. Then, from the properties of the $\chi^2$-distribution, the FAR can be written as,
\begin{align}
	\varkappa = 1-\frac{\gamma (\frac{\psi_{i,k}^T }{2},\frac{T}{2})}{\Gamma(\frac{T}{2})}, \label{eq_p_Tmu0}
\end{align}
where $\gamma(\cdot,\cdot)$ denotes the \textit{lower incomplete gamma function}.
Stateful detection can be extended to the case where more weights are considered for the recent residuals and less on the far past residuals. For this, define \textit{weighted} distance measure as sum of Chi-squared distributions \cite{Bausch2013OnTE,umsonst2019tuning},
\begin{equation} \label{eq_z2}
	\overline{\psi}_{i,k}^T = \sum_{m=k-T+1}^k \lambda^{k-m} \frac{(\mb{r}_m^i)^2}{\Phi}.
\end{equation}
where $0< \lambda \leq 1$ denotes the weight factor. Similar to Eq. \eqref{eq_theta_T}, the thresholds for this case are defined as, 
\begin{align} \label{eq_theta_T2}
	\mathbb{E}(\overline{\psi}_{i,k}^T) &= \frac{1-\lambda^{T}}{1-\lambda},\\ \label{eq_theta_T22} \theta_\varkappa^{T,\lambda} &= 2\Gamma^{-1}\left(1-\varkappa,\frac{1-\lambda^{T}}{2-2\lambda}\right).
\end{align}
Then, the FAR can be written as
\begin{align}
	\varkappa = 1-\frac{\gamma (\frac{\overline{\psi}_{i,k}^T}{2},\frac{1-\lambda^{T}}{2-2\lambda})}{\Gamma(\frac{1-\lambda^{T}}{2-2\lambda})}. \label{eq_p_Tmu}
\end{align}
With these formulations, the FDI strategy follows the logic in Algorithm~\ref{alg_logic2}.  
It should be noted that longer time-window $T$ (or larger weight $\lambda$) gives lower detection FAR while adding more \textit{delay} for raising the fault alarm. This implies a trade-off between the accuracy of the detection and the additive delay for declaring the alarm.

\begin{algorithm} [t] \label{alg_logic2}
	\textbf{Input:} $A$, $W$, $C$, $\mb{y}$, FAR $\varkappa$, $K$ from Algorithm~\ref{alg_lmi}, $T$,  $\lambda$ \\
	Find ${\psi}_{i,k}^T$ at CAV $i$ via \eqref{eq_z}, or $\overline{\psi}_{i,k}^T$ via \eqref{eq_z2}\;
	Define $\theta_\varkappa^T$ via \eqref{eq_theta_T}, or  $\theta_\varkappa^{T,\lambda}$ via \eqref{eq_theta_T2}\;
	\Begin(Stateful detection at CAV $i$:){
		\If{ ${\psi}_{i,k}^T \geq \theta_\varkappa^T$ or $\overline{\psi}_{i,k}^T \geq \theta_\varkappa^{T,\lambda}$}
		{Declare: fault alarm\;}
	}
	\textbf{Output} Binary decision (Alarm or no-Alarm) \;
	\caption{Stateful Residual-based FDI of NMVS}
\end{algorithm}

\section{Illustrative Simulations} \label{sec_sim}
\subsection{Distributed Observer Evaluation} \label{sec_sim_obsv}
For the first simulation, we consider the fault-free case to evaluate the performance of the proposed distributed observer \eqref{eq_p}-\eqref{eq_m}. Consider the NMVS in Fig.~\ref{fig_platoon} representing a mixed traffic setup of 4 CAVs and 4 HDVs. Every CAV only takes partial sensor measurement of only 1 HDV, with no direct measurement from other HDVs. To track the state of all HDVs, the CAVs share their measurement and a-priori estimates over the given strongly-connected communication network $\mc{G}$ and perform the consensus-based data fusion strategy given by distributed observer \eqref{eq_p}-\eqref{eq_m}.

Two well-known models are considered to simulate the HDVs' dynamics: (i) the so-called free-flow dynamics \cite{ahmed1999modeling} or (ii) Helly's car-following dynamics \cite{brackstone1999car}.  The free-flow model is used for the lead HDVs in the line of traffic, for example, HDV 3 and 4 in Fig.~\ref{fig_platoon}. This discrete-time model is given by:
\begin{align}  \label{eq_free-flow}
	v(k+1) = v(k) + \varrho (v_d(k) - v(k-\tau)) + \sigma(k),
\end{align}
with $v(k)$ denoting the velocity of the HDV at time $k$, $\tau$ as the reaction time factor, $\varrho$ as the coefficient related to how fast HDV follows the desired velocity $v_d(k)$, and the zero-mean Gaussian noise term $\sigma(k) \sim \mc{N}(0,\Sigma)$. 

For other HDVs, i.e., HDV 1 and 2 in Fig.~\ref{fig_platoon}, the car-following dynamics is used with the following discrete-time model \cite{brackstone1999car}:
\begin{align}  \label{eq_helly}
	v(k+1) = v(k) + a_1 \delta v(k-\tau)+ a_2 (\delta x(k-\tau) -D(k)),
\end{align}	
with $\delta v(k)$ and $\delta x(k)$ respectively denoting the difference of the velocity and position of the HDV and its front vehicle, $D(k) = b_1 + b_2 v(k-\tau)$ as the desired distance with $b_1$ and $b_2$ as the distance headway coefficients, $a_1$ and $a_2$ as some constants, and $\tau$ as the reaction time constant. 	The simulation parameters are given in Table~\ref{tab_sim}.
\begin{table} [h] 
	\centering
	\caption{The parameters chosen for the HDVs' dynamics. }
	\label{tab_sim}
	\begin{tabular}{|c|c|c|c|c|c|} 
		\hline
		$\varrho$ & $0.2$  
		&
		$\tau$ &  $10$ &   		
		$a_1$ &  $0.4$ \\   
		\hline
		$a_2$ &  $0.1$&
		$b_1$ &  $10$  &  
		$b_2$ &  $0.5$
		\\
		\hline
		\hline
	\end{tabular}
\end{table}	  

\begin{figure} 
	\centering
	\includegraphics[width=5in]{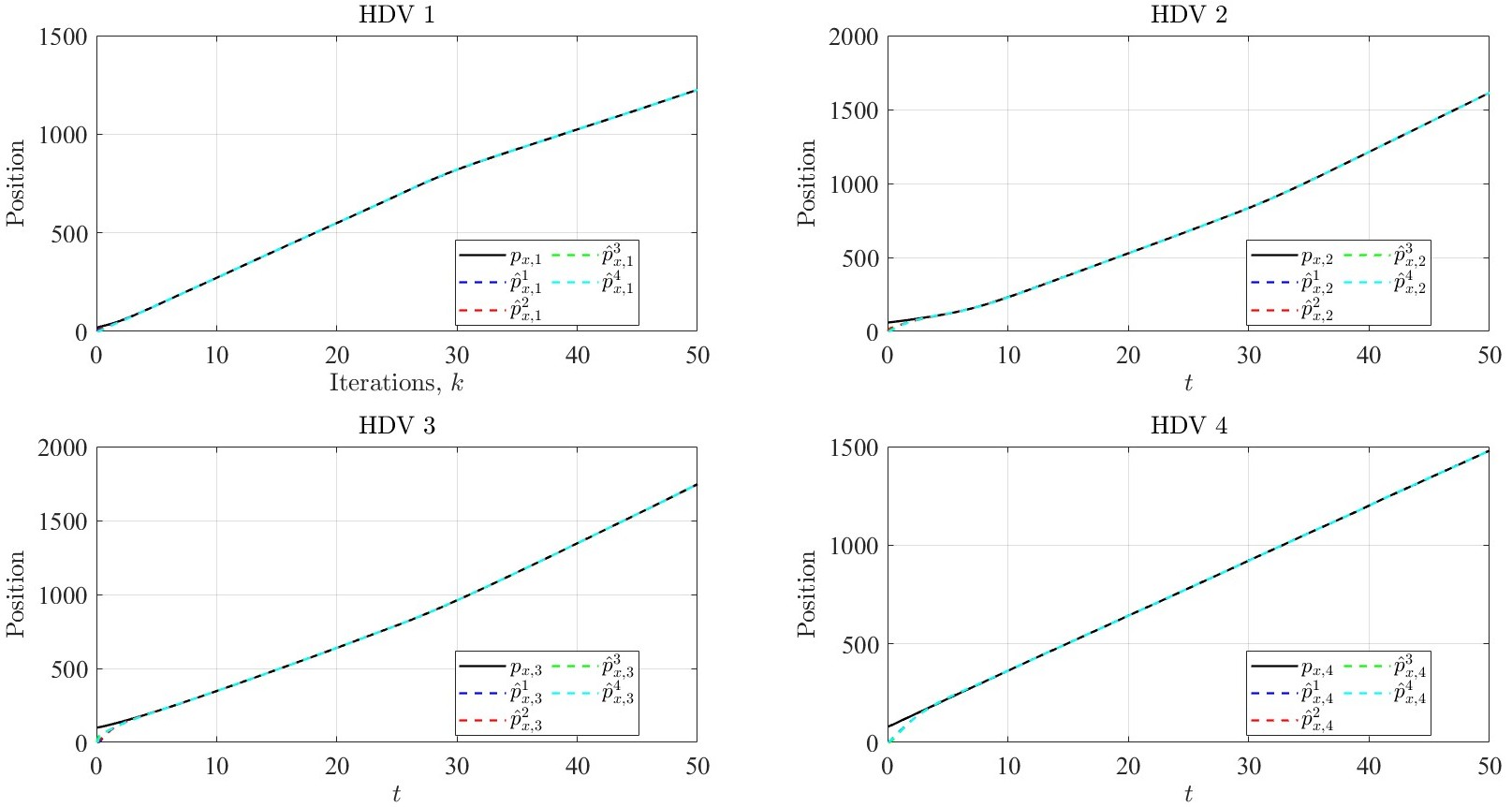}
	\caption{This figure presents the position state of the $i$-th HDV, denoted by $p_{x,i}$, for $i=1,\dots,4$ and the estimated position, $\hat{p}^j_{x,i}$ by the $j$-th CAV $j=1,\dots,4$ via the distributed observer \eqref{eq_p}-\eqref{eq_m}. 
	} \label{fig_mse1}
\end{figure}

\begin{figure} 
	\centering
	\includegraphics[width=5in]{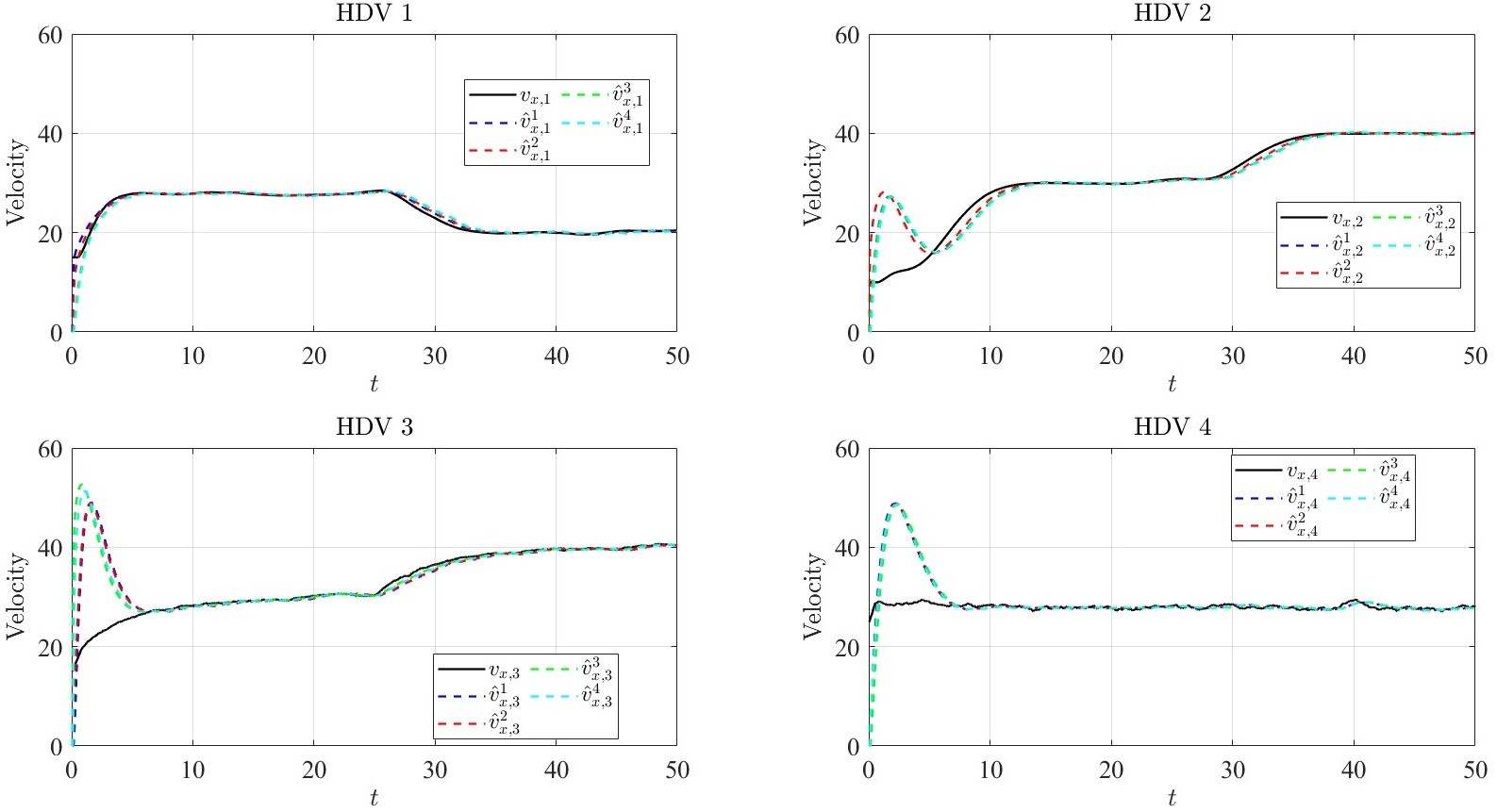}
	\caption{This figure presents the velocity state of the $i$-th HDV, denoted by $v_{x,i}$, for $i=1,\dots,4$ and the estimated velocity, $\hat{v}^j_{x,i}$ by the $j$-th CAV $j=1,\dots,4$ via the distributed observer \eqref{eq_p}-\eqref{eq_m}.
	} \label{fig_mse2}
\end{figure}

Following Remark~\ref{rem_ncv}, since the dynamics of HDVs are unknown for CAVs, each CAV assumes a nearly-constant-velocity (NCV) model for the $i$-th HDV dynamics $\mb{x}_{i,k+1}=\widetilde{A}_i\mb{x}_{i,k}+\nu_{i,k}$ with $\mb{x}_{i}=(p_{i,x};v_{i,x})$ (position and velocity in the $x$ direction) and
\begin{equation}
	\widetilde{A}_i=\left(
	\begin{array}{cc}
		1 & \mc{T} \\
		0 & 1 	
	\end{array} \right),
\end{equation}
with $\mc{T}$ as sampling time constant.
The consensus weight entries of $W$ are simply set as $\frac{1}{|\mc{N}_i|} = \frac{1}{3}$ with $\mc{N}_i$ including two neighboring CAVs and CAV $i$ itself. The CAVs share the direct measurement of velocity and position of 1 HDV and the a-priori estimates of velocity and position of all HDVs (i.e., $\widehat{\mb{x}}^j_{k-1|k-1}$). 
The measurement noise is $\mu_{i,k} \sim \mc{N}(0,0.15)$ and the velocity noise is $\sigma(k) \sim \mc{N}(0,0.1)$. The simulations are shown in time defined by $t=k\mc{T}$ with sampling time-step $\mc{T}=0.05 \rm{sec}$. HDV 3 changes its desired velocity at $t=25 \rm{sec}$ to $40 \rm{m/s}$. The simulation results on the positions and velocities of HDVs and the estimated velocities/positions by all 4 CAVs are shown in Fig.~\ref{fig_mse1}  and Fig.~\ref{fig_mse2}. As can be seen, all CAVs properly track and reach consensus on the velocity/position of HDVs in steady-state. There are minor steady-state errors due to uncertainties such as $\sigma(k)$ and other noise terms.

\subsection{Comparison over Large-Scale Setup}
\begin{figure} 
	\centering
	\includegraphics[width=2.5in]{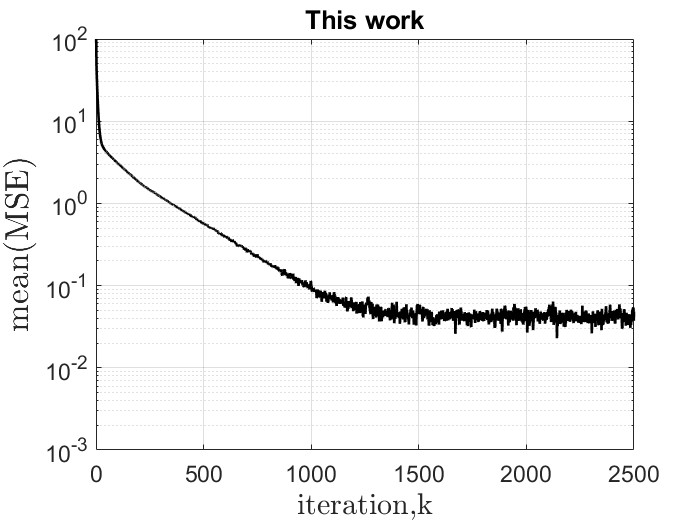}
	\includegraphics[width=2.5in]{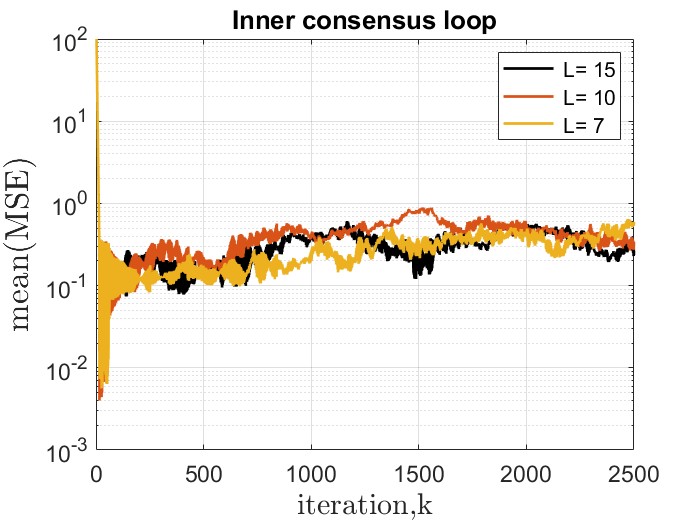}
	\caption{This figure compares the MSE performance of proposed distributed estimator \eqref{eq_p}-\eqref{eq_m} with the distributed estimator in \cite{he2020secure} having inner consensus loop.
	} \label{fig_mse_large}
\end{figure}

Next, over similar setup as in Section~\ref{sec_sim_obsv}, we consider a large-scale example vehicular network example (following Remarks~\ref{rem_comunic} and~\ref{rem_comp} on the scalability of the proposed observer design). We consider a mixed network of $50$ vehicles, including $n=25$ CAVs and $N=25$ HDVs. The network of CAVs is considered as an Erdos-Renyi graph with linking probability of $15\%$ with graph diameter $6$. The same NCV model with $\mc{T}=0.1sec$ is considered for the HDV model. On the other hand, the exact dynamics of odd-numbered HDVs follow the free-flow model defined by Eq.~\eqref{eq_free-flow} and the even-numbered HDVs follow the car-following dynamics given by Eq.~\eqref{eq_helly} with the same parameters as in Table~\ref{tab_sim}. We apply the proposed distributed observer to track the state of HDVs by the CAVs. The mean-square-error (MSE) of the state estimation is averaged at all CAVs, and the results are shown in Fig.~\ref{fig_mse_large}. For comparison, the MSE performance is compared with the distributed estimator in \cite{he2020secure}, which is designed based on inner consensus loop strategy, i.e., it performs $L$ steps of consensus iterations between every two consecutive time steps of system dynamics $k$ and $k+1$. This inner consensus loop parameter $L$ should be larger than the network diameter (we set $L=7,10,15$). This implies that the proposed strategy in \cite{he2020secure} needs $L$ time faster communication and processing units at CAVs, with $L$ times more computation and communication overhead. Therefore, as a comparison, our proposed strategy is computationally/communicationally more efficient, while giving better MSE performance as shown in Fig.~\ref{fig_mse_large}.  

\subsection{Distributed FDI Evaluation}
\begin{figure} 
	\centering
	\includegraphics[width=3in]{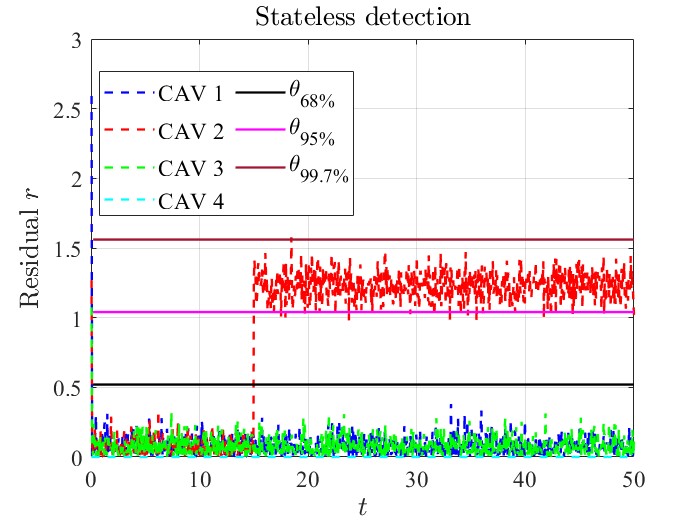}
	\caption{This figure presents the residuals $\mb{r}_k^i$, for $i=1,\dots,4$ under the distributed observer \eqref{eq_p}-\eqref{eq_m}. The faulty measurement at CAV 2 is detected by FAR less than $5\%$ via the proposed stateless FDI.
	} \label{fig_r}
\end{figure}

Next, we evaluate the performance of the proposed distributed FDI strategy. Assume that sensor measurement by CAV 2 in Fig.~\ref{fig_platoon} is faulty, i.e., the measured state of HDV 2 is erroneous or biased due to adversarial conditions or possible attacks. To model this, we consider $f_{2,k}\sim \mc{N}(1.5,0.25)$ for $k\geq 300$ (or $t\geq 15$). First, we check the instantaneous residuals $\mb{r}_k^i$ under the proposed distributed observer \eqref{eq_p}-\eqref{eq_m} for $i=1,\dots,4$. We find the probabilistic thresholds $\theta_\kappa$ via Eq.~\eqref{eq_thresold} for three detection probabilities. For the given observer and system setup in this example, we have $\Phi = 0.521$. The simulation results are shown in Fig.~\ref{fig_r}. Clearly, in steady-state, the residuals $\mb{r}_k^i$ for $i=1,3,4$ are below the given thresholds, while $\mb{r}_k^2$ is above $\theta_{95\%}$ for $t\geq 15$ (this implies FAR less than $5\%$). Therefore, the faulty measurement is isolated at CAV 2 and is not cascaded to other CAVs because of information sharing.

\begin{figure} 
	\centering
	\includegraphics[width=3in]{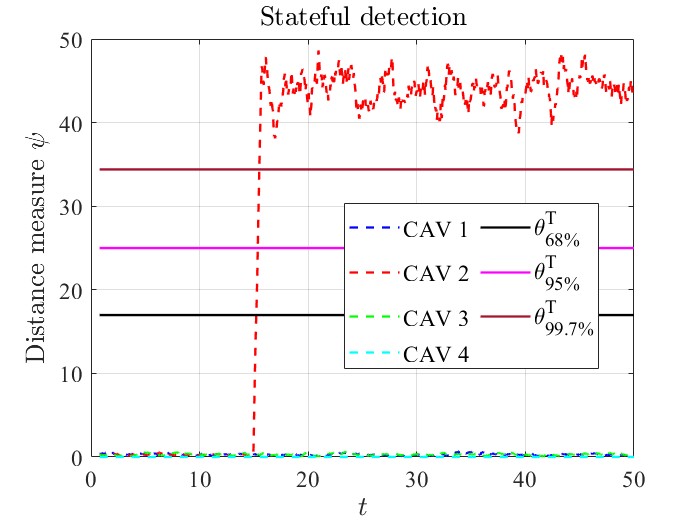}
	\caption{This figure presents the distance measures $\psi_{i,k}^T$, for $i=1,\dots,4$ and $T=15$ time-steps (or $0.75\rm{sec}$). The faulty measurement at CAV 2 is detected by FAR less than $0.3\%$ via the proposed stateful FDI.
	} \label{fig_psi}
\end{figure}

For the same setup, we evaluate the performance of our stateful detection strategy over a sliding time-window of length $T=15$ time-steps (or $0.75\rm{sec}$). The distance measures $\psi_{i,k}^T$ for $i=1,\dots,4$ are defined via \eqref{eq_z} and the thresholds $\theta_\varkappa^T$ are designed based on \eqref{eq_theta_T} for three FARs. The simulation results are shown in Fig.~\ref{fig_psi}. Clearly, the distance measures $\psi_{i,k}^T$ for $i=1,3,4$ are below the probabilistic thresholds, while $\psi_{2,k}^T$ is above $\theta^T_{99.7\%}$ for $t\geq 15$. For this case, the FAR is less than $0.3\%$. This shows that this stateful FDI scenario is less prone to false alarms. However, this comes with extra complexity and more memory usage for detection.

\begin{figure} 
	\centering
	\includegraphics[width=3in]{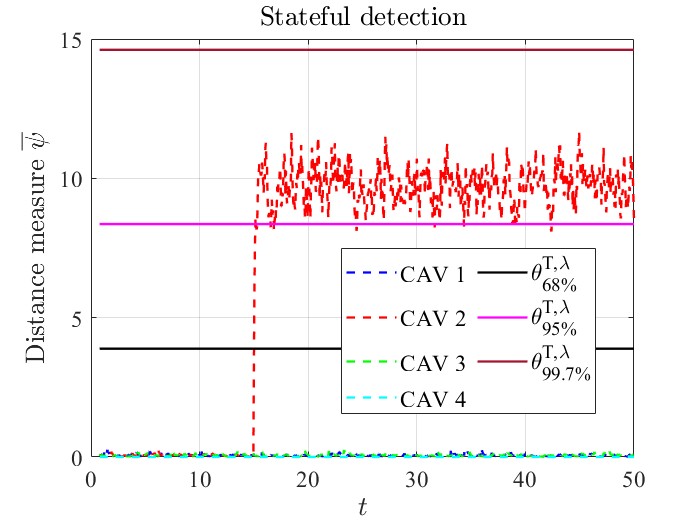}
	\caption{This figure presents the weighted distance measures $\overline{\psi}_{i,k}^T$, for $i=1,\dots,4$, $T=15$ time-steps (or $0.75\rm{sec}$), and weight factor $\lambda=0.7$. The faulty measurement at CAV 2 is detected by FAR less than $5\%$ via the proposed stateful FDI.
	} \label{fig_psi2}
\end{figure}

\begin{figure} 
	\centering
	\includegraphics[width=3in]{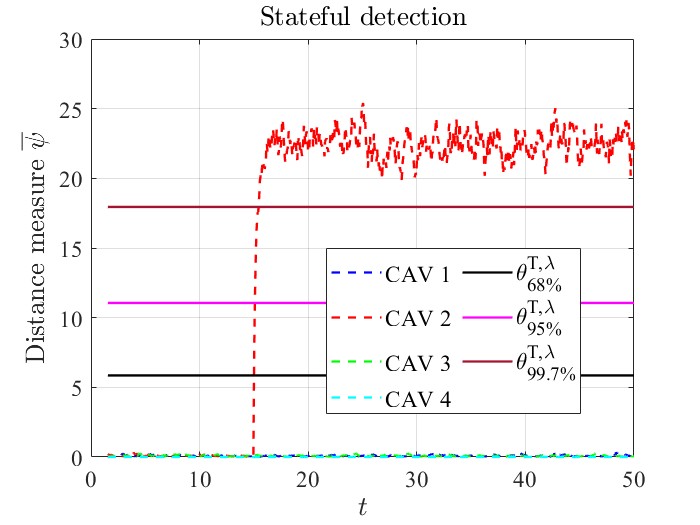}
	\caption{This figure presents the weighted distance measures $\overline{\psi}_{i,k}^T$, for $i=1,\dots,4$, $T=30$ time-steps (or $1.5\rm{sec}$), and weight factor $\lambda=0.8$. The faulty measurement at CAV 2 is detected by FAR less than $0.3\%$ via the proposed stateful FDI.
	} \label{fig_psi3}
\end{figure}

Similarly, the weighted distance measures can be considered for stateful detection. For this, we consider the weight factor $\lambda=0.7$. The weighted distance measures $\overline{\psi}_{i,k}^T$ for $i=1,\dots,4$ and the associated thresholds $\theta_\varkappa^{T,\lambda}$ are defined based on \eqref{eq_z2} and \eqref{eq_theta_T22}. The simulation results are given in Fig.~\ref{fig_psi2}. It is clear from the figure that only the distance measure $\overline{\psi}_{2,k}^T$ is above $\theta^{T,\lambda}_{95\%}$ for $t \geq 15$. For this case, the FAR is less than $5\%$. 
One can reduce the FAR by increasing the length of the sliding time-window and/or increasing the weight factor. We redid this simulation for $T=30$ time-steps (or $1.5\rm{sec}$) and $\lambda=0.8$. The results are shown in Fig.~\ref{fig_psi3}. For this case, the FAR is less than $0.3\%$. However, this lower FAR comes with more computational complexity, data storage, and some minor delay in detection. 
Note that, for both stateful cases, the fault is clearly isolated without cascading effect on the rest of the CAVs.

\subsection{FDI under Large Noise/Uncertainty}
\begin{figure} 
	\centering
	\includegraphics[width=5in]{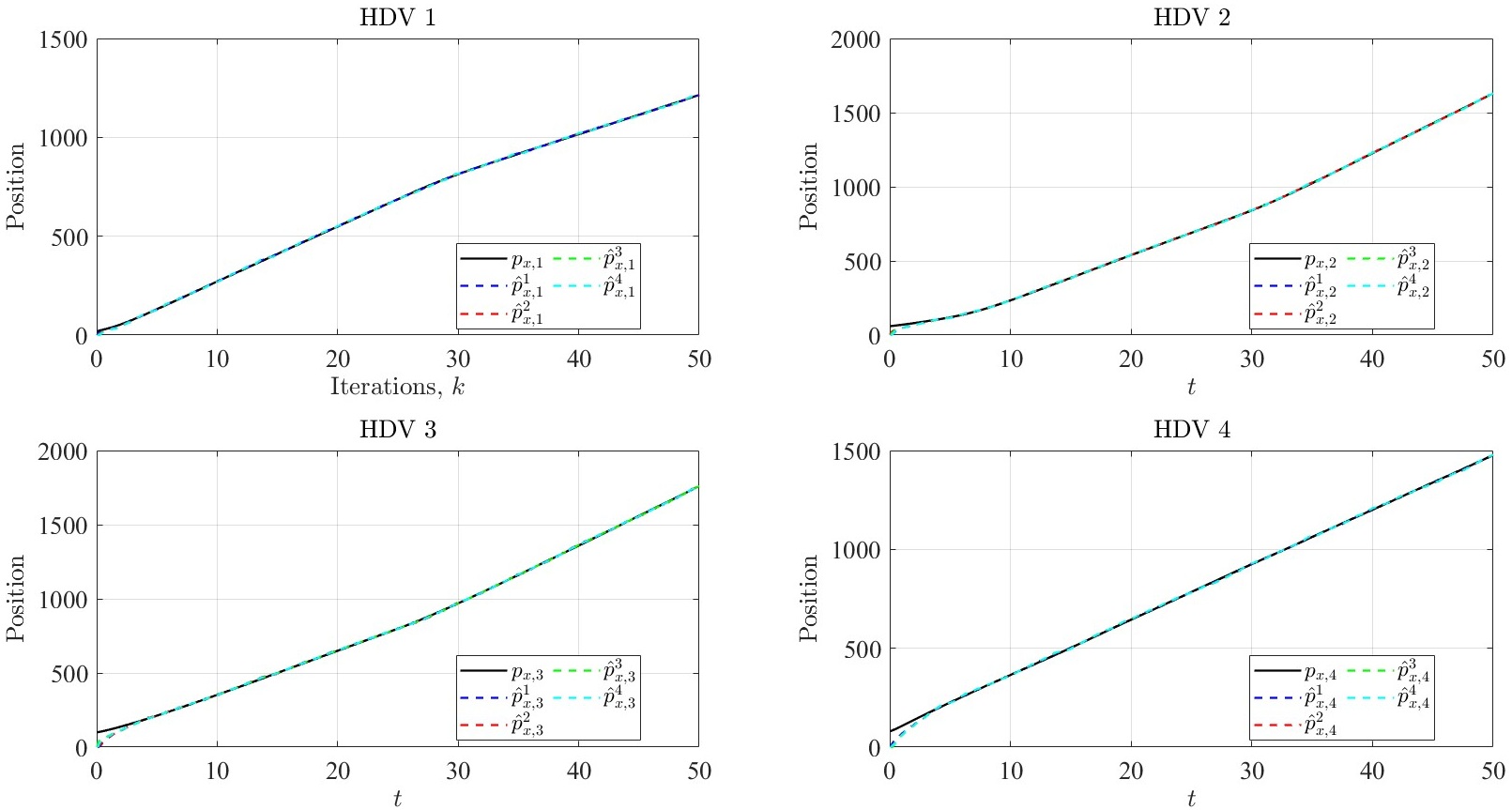}
	\caption{This figure shows the position of the $i$-th HDV, denoted by $p_{x,i}$, for $i=1,\dots,4$ and the estimated position, $\hat{p}^j_{x,i}$ by the $j$-th CAV $j=1,\dots,4$ via the distributed observer \eqref{eq_p}-\eqref{eq_m}.  
	} \label{fig_mse1_noise}
\end{figure}

\begin{figure} 
	\centering
	\includegraphics[width=5in]{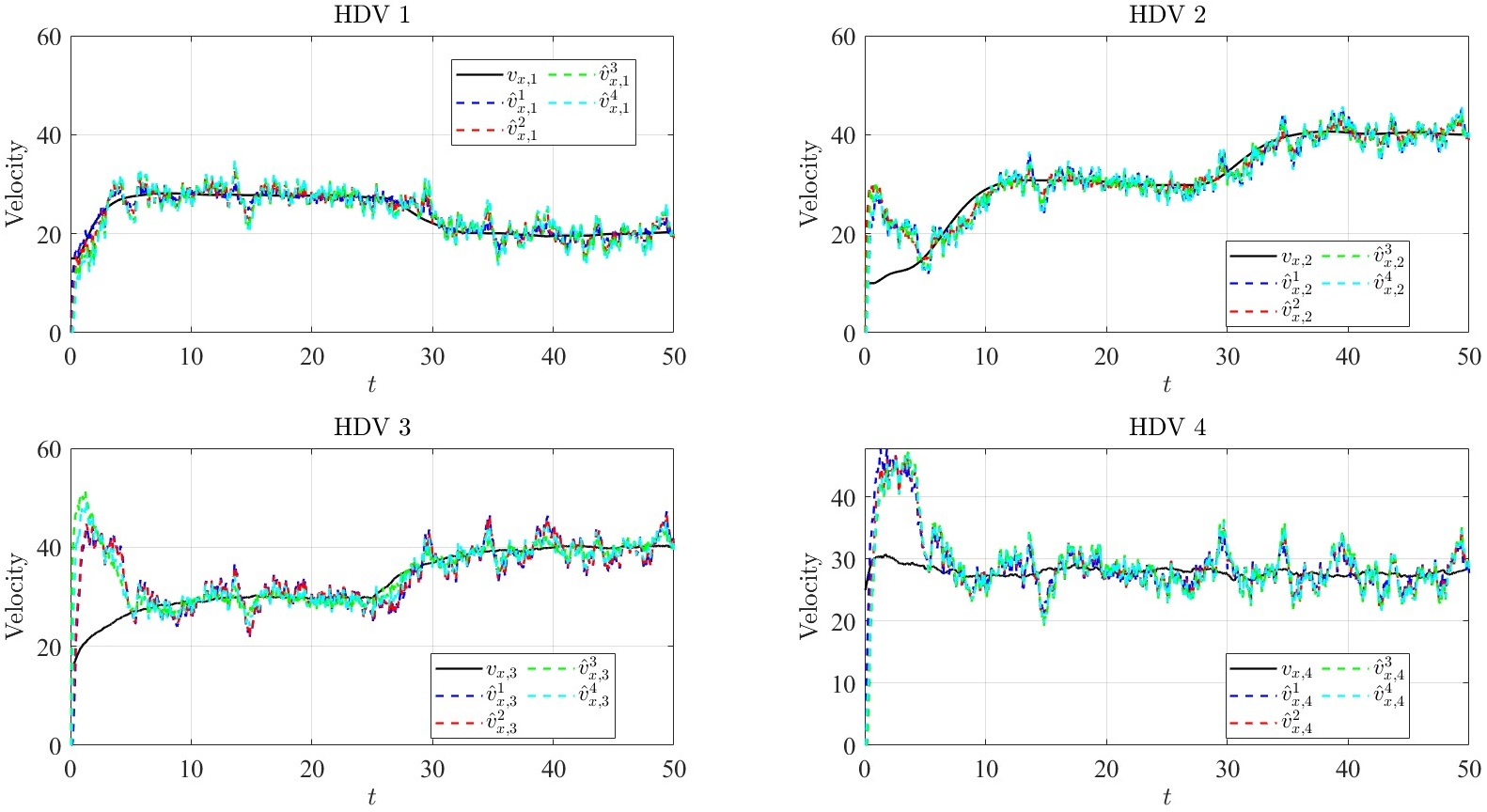}
	\caption{This figure shows the velocity of the $i$-th HDV, denoted by $v_{x,i}$, for $i=1,\dots,4$ and the estimated velocity, $\hat{v}^j_{x,i}$ by the $j$-th CAV $j=1,\dots,4$ via the distributed observer \eqref{eq_p}-\eqref{eq_m}.
	} \label{fig_mse2_noise}
\end{figure}

Next, we evaluate our FDI technique subject to large noise and uncertainty. For this simulation, we consider large Gaussian noise term $\sigma(k) \sim \mc{N}(0,5)$ in free-flow model given by Eq.~\eqref{eq_free-flow}. For the system of HDVs in Eq.~\eqref{eq_sys1} we set large noise $\nu_k \sim \mc{N}(0,1)$ and for the measurements in Eq.~\eqref{eq_H} we set large noise $\mu_k \sim \mc{N}(0,1)$. By this setup, we simulate the mixed traffic tracking and FDI scenario subject to large uncertainty. The rest of the simulation parameters are set the same as in previous subsections. We first evaluate the observer performance and the simulation results are shown in Fig.~\ref{fig_mse1_noise} and \ref{fig_mse2_noise}. It is clear from the figures that the CAVs are able to track the state of HDVs despite large uncertainty. It should be clarified that the HDV parameters by free-flow and Helly's car-following models are generally unknown to the CAVs, and CAVs only track the HDVs based on the measurements shared over the communication network using simplified NCV/NCA model as the system dynamics, see Remark~\ref{rem_ncv}. In other words, the only parameter used in the observer design is sampling time $\mc{T}$ along with the noisy measurements.

\begin{figure} 
	\centering
	\includegraphics[width=3in]{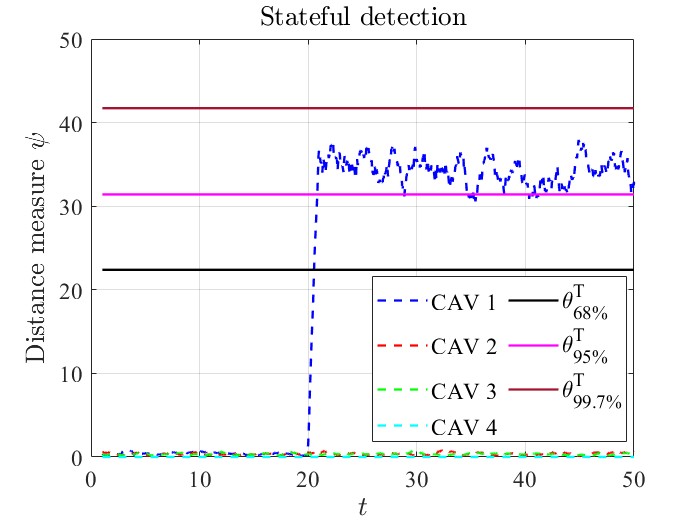}
	\caption{This figure shows the distance measures $\psi_{i,k}^T$, for $i=1,\dots,4$ and $T=20$ time-steps (or $1\rm{sec}$). The fault at CAV 1 is detected by FAR less than $5\%$ via the proposed stateful FDI.
	} \label{fig_psi_noise}
\end{figure}

\begin{figure} 
	\centering
	\includegraphics[width=3in]{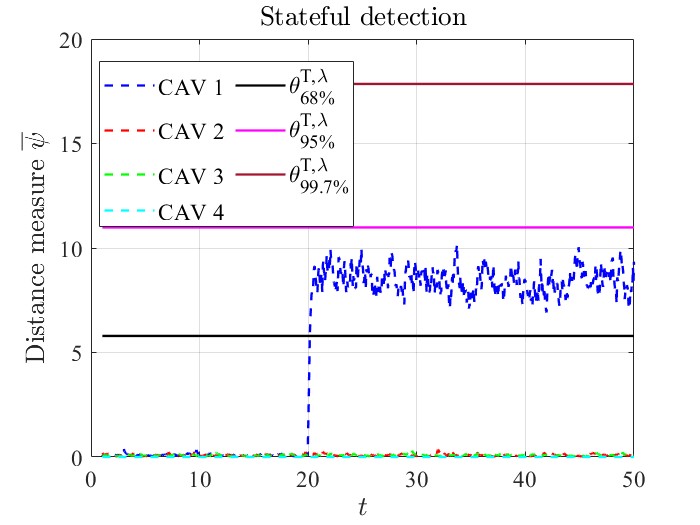}
	\caption{This figure shows the weighted distance measures $\overline{\psi}_{i,k}^T$, for $i=1,\dots,4$, $T=20$ time-steps (or $1\rm{sec}$), and weight factor $\lambda=0.8$. The fault at CAV 1 is detected by FAR less than $32\%$.
	} \label{fig_psi_noise2}
\end{figure}

Next, we consider possible fault at CAV 1 as $f_{1,k}\sim \mc{N}(5,0.5)$ for $k\geq 400$. We consider stateful detection over time-window $T=20$ or $1\rm{sec}$. The measures $\psi_{i,k}^T$ for $i=1,\dots,4$ are based on Eq. \eqref{eq_z} and the thresholds $\theta_\varkappa^T$ are based on \eqref{eq_theta_T} for three FARs. Due to large noise values, the detection thresholds are also large. This imlpies that the FDI results in high FAR. The simulation results are given in Fig.~\ref{fig_psi_noise}, where  $\psi_{1,k}^T$ is above $\theta^T_{95\%}$ for $t\geq 20$. For this case, the FAR is less than $5\%$. 
We also consider weighted distance measures $\overline{\psi}_{i,k}^T$ for $i=1,\dots,4$ with weight factor $\lambda=0.8$. The associated thresholds $\theta_\varkappa^{T,\lambda}$ are based on \eqref{eq_z2} and \eqref{eq_theta_T22}. Due to large noise values, the detection thresholds are also large, resulting in higher FAR. The simulation results are shown in Fig.~\ref{fig_psi_noise2}, where the distance measure $\overline{\psi}_{1,k}^T$ is above $\theta^{T,\lambda}_{68\%}$ for $t \geq 20$. For this case, the FAR is less than $32\%$. 

\section{Conclusions}\label{sec_con}
\subsection{Concluding Remarks}
A distributed FDI approach in mixed traffic networks involving CAVs and HDVs with partial sensor measurements is presented. The proposed consensus-based observer is shown to enable each CAV to estimate the state of all HDVs without assuming local observability. The proposed local residual-based FDI schemes developed here (both stateful and stateless) further allow each CAV to independently detect and isolate faults or attacks on its sensors via probabilistic threshold design accounting for realistic noise characteristics. The proposed methods demonstrate the potential of decentralized fault detection to improve reliability and safety in heterogeneous traffic setups without relying on centralized processing.

\subsection{Future Directions}
In large-scale setups, the communication network of vehicles may encounter latency. Designing \textit{delay-tolerant} distributed observers is one direction of our future research.
One can extend the results for tracking of platooning with heterogeneous vehicle fleets. Resilient and optimal design of the CAV network via survivable network design is another interesting research direction.	
The results can also be extended to observer-based tracking of different networked systems, e.g., for distributed target tracking and localization. Integrating cooperative control strategies with the proposed FDI methods is another future research direction to enable fault-tolerant decision-making and coordinated path planning.  

\bibliographystyle{elsarticle-num}
\bibliography{bibliography}

\end{document}